\documentclass[%
 reprint,
 amsmath,amssymb,
 aps,
prstab,
]{revtex4-2}

\usepackage{graphicx}
\usepackage{dcolumn}
\usepackage{bm}

\begin{document}

\preprint{APS/123-QED}

\title{Minimizing the fluctuation of resonance driving terms in dynamic aperture optimization}

\author{Bingfeng Wei$^1$}

\author{Zhenghe Bai$^{1,2}$}
\email{Corresponding author. baizhe@ustc.edu.cn}

\author{Jiajie Tan$^1$}

\author{Lin Wang$^1$}

\author{Guangyao Feng$^1$}%
\email{Corresponding author. fenggy@ustc.edu.cn}

\affiliation{
    $^1$National Synchrotron Radiation Laboratory, University of Science and Technology of China, Hefei 230029, China
    }

\affiliation{
    $^2$Synchrotron SOLEIL, L'Orme des Merisiers, Saint-Aubin, France
    }

\date{\today}

\begin{abstract}
    Dynamic aperture (DA) is an important nonlinear property of a storage ring lattice,
    which has a dominant effect on beam injection efficiency and beam lifetime.
    Generally, minimizing both resonance driving terms (RDTs) and amplitude dependent tune shifts is an essential condition for enlarging the DA.
    In this paper, we study the correlation between the fluctuation of RDTs along the ring and the DA area with double- and multi-bend achromat lattices.
    It is found that minimizing the RDT fluctuations is more effective than minimizing RDTs themselves in enlarging the DA, and thus can serve as a very powerful indicator in the DA optimization.
    Besides, it is found that minimizing lower-order RDT fluctuations can also reduce higher-order RDTs, which are not only more computationally complicated but also more numerous.
    The effectiveness of controlling the RDT fluctuations in enlarging the DA confirms that the local cancellation of nonlinear effects used in some diffraction-limited storage ring lattices is more effective than the global cancellation.
\end{abstract}

\maketitle


\section{\label{sec:level1}Introduction}
Dynamic aperture (DA) has a dominant effect on beam injection efficiency and beam lifetime of a storage ring. 
Optimization of the DA is a complex problem with a long history. In the past decade or so, due to the improvement of computer performance and the application of evolutionary algorithms, particle tracking-based numerical approach has been widely used for DA optimization, in which genetic algorithm or particle swarm optimization algorithm is applied to find the globally best solution \cite{Borland2009,Yang2011, Gao2011, Bai2011,HUANG201448,Ehrlichman2016}.
But this numerical approach is quite demanding in computational resources, and in general, there is basically no physics to guide further lattice optimization.
As an alternative and complementary approach, resonance driving term (RDT) minimization \cite{SLS97} is a traditional analytical approach with fast optimization speed and easily-revealed physics. In this analytical approach, minimizing RDTs to suppress the corresponding resonances and also controlling amplitude dependent tune shifts (ADTS) to avoid resonance crossings can help to enlarge the DA. However, small RDTs is a necessary but not sufficient condition for large DA \cite{Yang2011}, and the optimization result obtained by this approach largely depends on the lattice designers' experiences. 

Nevertheless, the guidance of the RDTs is of great significance. Two types of nonlinear cancellation schemes, which are made within one lattice cell, were proposed in the multi-bend
achromat (MBA) lattice
design of diffraction-limited storage rings (DLSRs) and showed remarkable success \cite{ESRFEBS, sls2}. One is the hybrid MBA lattice with a pair of -$\mathcal{I}$ separated dispersion bumps \cite{ESRFEBS}, and the other is the higher-order achromat (HOA) lattice with some identical bend unit cells \cite{sls2}. Both can cancel the main RDTs generated by sextupoles within one lattice cell.
This local cancellation prevents the RDTs from building up along 
the ring and is thus more effective than the global cancellation made over some lattice cells \cite{book2020}. 
Moreover, minimizing the turn-by-turn fluctuations of the Courant-Snyder actions for particles helps to enlarge the DA \cite{Li2021}.
The Courant-Snyder action fluctuations could be related to the RDT fluctuations.
This inspires us the importance of suppressing the building up of RDTs,
or in other words, minimizing the RDT fluctuations along the ring.
In this paper the RDTs will be calculated as a function of the position along the ring.
We will step further to consider their fluctuations, 
and try to find the correlation between the DA area and the RDT fluctuations with a large number of nonlinear lattice solutions.

The remaining sections of this paper are outlined as follows. 
Section \ref{sec:II} introduces the RDTs briefly and describes their fluctuations along the ring.
Then, in Section \ref{sec:III}, the study starts with the simple double-bend achromat (DBA) lattice of a third-generation synchrotron light source, where low-order RDTs are the most important.
Next we step further into the more complex case of two 6BA lattices of DLSRs in Section \ref{sec:IV}. At the end of the paper, a brief summary and outlook are given.

\section{\label{sec:II}control of the RDT fluctuations} 
The one-turn map of a storage ring with $N+1$ linear maps separated by $N$ thin-lens sextupole maps can be normalized as \cite{SLS97}:
\begin{eqnarray} \label{eq1}
    \mathcal{M}_{1\rightarrow N+1} &=& \mathcal{M}_{1\rightarrow 1} e^{:V_1:}\mathcal{M}_{1\rightarrow 2} e^{:V_2:} ...  
        e^{:V_{N}:}\mathcal{M}_{N\rightarrow N+1} \nonumber\\
        &=& \mathcal{A}_0^{-1} e^{:\hat{V}_1:}e^{:\hat{V}_2:} ... e^{:\hat{V}_{N}:} \mathcal{R}_{1\rightarrow N+1}\mathcal{A}_{N+1} \nonumber \\
        &=& \mathcal{A}_0^{-1} e^{:h:}  \mathcal{R}_{1\rightarrow N+1}\mathcal{A}_{N+1},
\end{eqnarray}
where $\hat{V}_i \equiv \mathcal{R}_{1\rightarrow i} \mathcal{A}_i V_i$, $\mathcal{A}$ is a normalizing map, $\mathcal{R}$ is a rotation, $e^{:h:}$ is the nonlinear Lie map.
Using the resonance basis, the $n$-th order generator of $e^{:h:}$ can be expanded as:
\begin{eqnarray} \label{RDT}
    h_n = \sum_{n=j+k+l+m+p} h_{jklmp} h_x^{+j} h_x^{-k} h_y^{+l} h_y^{-m} \delta^{p} ,
\end{eqnarray}
where $h_x^{\pm}\equiv\sqrt{2 J_x}e^{\pm i\phi_x}$, $h_y^{\pm}\equiv\sqrt{2 J_y}e^{\pm i\phi_y}$, with $(J, \phi)$ being action-angle variables, and $h_{jklmp}$ is the so-called driving terms. The terms with $p\neq 0$ are chromatic terms, which affect the off-momentum dynamics.
In this paper we focus on the on-momentum DA, where the geometric terms with $p = 0$ are considered.
The geometric terms can be divided into two categories.
The terms with $j=k$ and $l=m$ drive the ADTS, and the remaining terms drive resonances $(j-k)\nu_x + (l-m)\nu_y$.

The concept of RDTs is derived from the one-turn map, and traditionally, one focuses on the RDTs of a periodic map or one-turn map.
In this paper we take the fluctuation of RDTs along the ring into consideration.
Denoting $\prod_{a=1}^t e^{:\hat{V}_a:} \equiv e^{:S_t:}$, $e^{:S_t:} = e^{:S_{t-1}:}e^{:\hat{V}_t:}$,  and when $t=N$, $S_N$ is the $h$ of Eq.~(\ref{eq1}).
According to the Baker-Campbell-Hausdorff formula \cite{Chao_book}, we have
\begin{widetext}
\begin{eqnarray} \label{BCH}
    S_t = S_{t-1} + \hat{V}_t + \frac{1}{2} :S_{t-1}:\hat{V}_t +\frac{1}{12}:S_{t-1}:^2 \hat{V}_t+\frac{1}{12}:\hat{V}_t:^2 S_{t-1} + \dots .
\end{eqnarray}
\end{widetext}
Equation~(\ref{BCH}) indicates that the lower-order terms of $S_{t-1}$
contribute to the higher-order terms of $S_{t}$.
Expanding $S_{t}$ with the resonance basis as in Eq.~(\ref{RDT}), we can get a series of nonlinear terms that show the change (or fluctuation) of driving terms $h_{jklmp}$ along the ring. We denote the driving terms of $S_t$ as $h_{1\rightarrow t, jklmp}$, and then one-turn RDTs can be written as $h_{jklmp} = h_{1\rightarrow t, jklmp} + h_{t+1 \rightarrow N, jklmp}$.
The lower-order RDT fluctuations contribute to the higher-order RDTs of the one-turn map.
For example, the fourth-order RDTs of sextupoles are crossing terms of their third-order RDTs \cite{SLS97},
\begin{eqnarray} \label{crossing}
    h_4 = \frac{1}{2}\sum_{b>a=1}^{N} \left[\hat{V}_{a}, \hat{V}_{b}\right]= \frac{1}{2}\sum_{b=2}^{N}\left[\sum_{a=1}^{b-1} \hat{V}_a, \hat{V}_b\right],
\end{eqnarray}
where $\sum_{a=1}^{b-1} \hat{V}_a$ is the third-order term of $S_{b-1}$.
Reducing the amplitude of $h_{1\rightarrow t, jklmp}$ can be beneficial for controlling the crossing terms.
Moreover, the discussion above can also apply to a storage ring with both sextupoles and octupoles.
The crossing terms of sextupoles and octupoles contribute to the fifth- and higher-order RDTs.
Therefore, the process of the nonlinear driving term fluctuating along the ring can provide more dynamics information.

\begin{figure}
    \includegraphics[width=\linewidth]{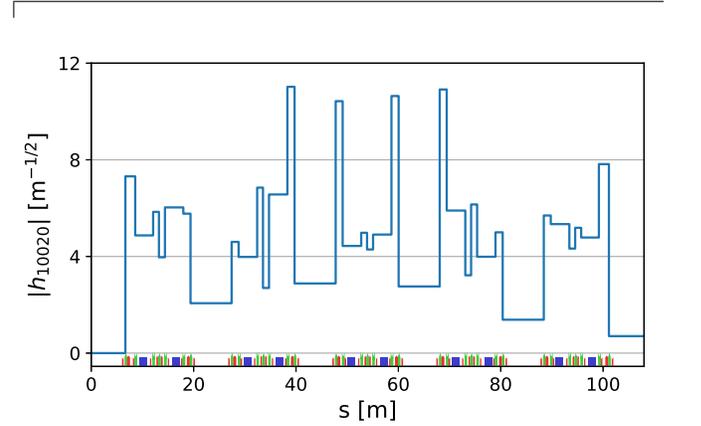}
    \caption{\label{fig:SSRF_RDTs_change}The change of the third-order RDT $h_{10020}$ at the locations of sextupoles (green blocks in the magnet layout) in one SP of the SSRF storage ring lattice.}
\end{figure}

In order to clearly illustrate the fluctuations of RDTs, the third-order RDT $h_{10020}$ of the SSRF storage ring lattice is plotted in Fig. \ref{fig:SSRF_RDTs_change} as a function of position in one super-period (SP).
It changes stepwise at the locations of sextupoles.
Traditionally, in order to enlarge the on-momentum DA, it is necessary to control the values of geometric RDTs of one-turn map, denoted as $h_{jklm0,\mathrm{ring}}$.
In this paper, we will control the average amplitudes of RDTs at all locations of nonlinear magnets along the ring, i.e. $\sum_{t=1}^N \left|h_{1\rightarrow t, jklm0}\right|/ N \equiv h_{jklm0, \mathrm{ave}}$.

\begin{figure}
        \includegraphics[width=0.8\linewidth]{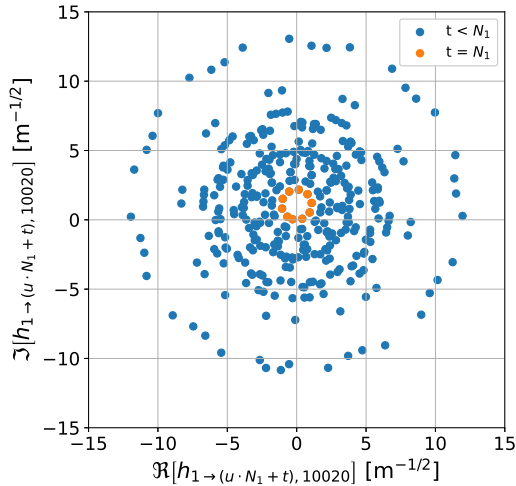}
        \caption{\label{fig:SSRF_complex_RDTs}Fluctuation of $h_{10020}$ of the SSRF lattice in the complex plane for 10 SPs.}
\end{figure}

In the complex plane, we can characterize the fluctuation of RDTs more clearly and show the regularity.
Referring the definition in Ref.~\cite{OPA}, we introduce $\bm{m} = (j - k, l-m)$ to represent the mode of $h_{jklm0}$ and $\bm{\mu} = 2 \pi (\nu_x, \nu_y)$ the phase advances of one SP.
We can use the RDT fluctuation data of one SP to construct the fluctuation over any number of SPs.
For each third-order RDT $h_{jklm0}$, if we denote the number of sextupoles of one SP as $N_1$,
the value of $h_{jklm0}$ at the $t$-th sextupole of the $(u+1)$-th SP ($1\leq t \leq N_1, u \geq 0)$ is
\begin{eqnarray} \label{third_RDT}
        & h_{1\rightarrow (u\cdot N_{1}+t), jklm0}\nonumber \\
        = &h_{1\rightarrow u\cdot N_{1}, jklm0}   + h_{\left(u\cdot N_{1}+1\right) \rightarrow (u\cdot N_{1} + t), jklm0}\nonumber \\
        = & h_{1\rightarrow N_{1}, jklm0} \frac{1 - e^{iu\bm{m}\cdot \bm{\mu}}}{1 - e^{i\bm{m}\cdot \bm{\mu}}} + h_{1\rightarrow t, jklm0} e^{iu\bm{m}\cdot \bm{\mu}}\nonumber \\
        = &  \frac{h_{1\rightarrow N_{1}, jklm0}}{1 - e^{i\bm{m}\cdot \bm{\mu}}} + \left(h_{1\rightarrow t, jklm0} - \frac{h_{1\rightarrow N_{1}, jklm0}}{1 - e^{i\bm{m}\cdot \bm{\mu}}}\right) e^{iu\bm{m}\cdot \bm{\mu}}.
\end{eqnarray}
The calculation of multi-period RDTs in Eq.~(\ref{third_RDT}) was derived in Ref.~\cite{CERN8805}.
With $u$ as a variable, Eq.~(\ref{third_RDT}) is described by a constant term and the $e^{iu\bm{m}\cdot \bm{\mu}}$ term, forming a circle with the center not at the origin in the complex plane.
Also taking the SSRF lattice as an example, we calculated the fluctuation of $h_{10020}$ for 10 SPs, and the results are plotted in the complex plane in Fig.~\ref{fig:SSRF_complex_RDTs}.
For $N_1$ sextupoles, there are $N_1$ concentric circles as the dots shown in Fig.~\ref{fig:SSRF_complex_RDTs}.
And when $t=N_1$, the circle passes the origin as shown with the orange dots.
The constant term, which is the center of these circles, shows the overall offset.
Traditionally, minimizing one RDT only involves the constant term.
While reducing fluctuation of one RDT involves reducing both the radii of these circles and the offset of the center.
The case of the fourth-order RDTs is more complex, shown in the appendix.

Different RDTs driving different nonlinear effects are not of the same importance.
It will be complicated to consider individual weights for different RDTs.
For simplicity, in this paper, the RDTs of the same order have the same weight.
The fluctuation of the $n$-th order RDTs, denoted as $h_{n, \mathrm{ave}}$, is calculated as
\begin{eqnarray} \label{h_n_ave}
    h_{{n,\mathrm{ave}}} = \sqrt{\sum_{j+k+l+m=n} (h_{j k l m 0, \mathrm{ave}})^2}.
\end{eqnarray}
And we use $h_{n, \mathrm{ring}}$ to represent the $n$-th order one-turn RDTs, which is defined in the same way as in Eq.~({\ref{h_n_ave}}).
When the third- and fourth-order RDTs are considered simultaneously, we introduce a weight coefficient $w$ for the fourth-order RDTs.
For example, the sum of the third- and fourth-RDT fluctuations is calculated as $h_{3, \mathrm{ave}} + w\cdot h_{4,\mathrm{ave}}$.
Besides, the ADTS terms also affect the on-momentum DA.
We denote the one-turn ADTS terms as $h_\mathrm{ADTS}$, which is calculated as
\begin{eqnarray}
    h_{\mathrm{ADTS}} = \sqrt{\left(\frac{d\nu_x}{dJ_x}\right)^2 + \left(\frac{d\nu_x}{dJ_y}\right)^2 + \left(\frac{d\nu_y}{dJ_y}\right)^2}.
\end{eqnarray}

By the way, calculating the RDT fluctuations is a necessary step to calculate the one-turn RDTs, which requires almost no additional calculations.
With the data of RDT fluctuations stored in the calculation, we can directly have the values of $\sum_{a=1}^{b-1} \hat{V}_a$ in Eq.~\ref{crossing}, allowing us to calculate the crossing terms using only one loop.

\section{\label{sec:III}Optimization of a DBA lattice}
\begin{figure}
    \includegraphics[width=\linewidth]{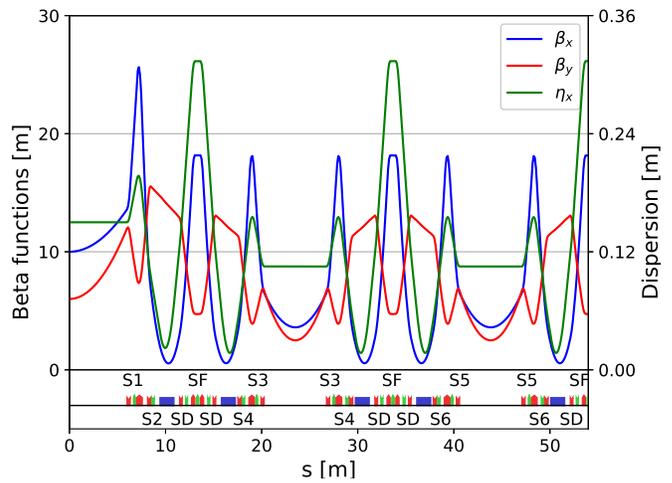}
    \caption{\label{fig:SSRF_optics}Linear optical functions and magnet layout of a half SP of the SSRF lattice.
    In the layout, bends are in blue, quadrupoles in red and sextupoles in green.}
\end{figure}

Now we first use the SSRF lattice to analyze the nonlinear dynamics based on RDTs and their fluctuations.
SSRF is a third-generation synchrotron light source with a beam energy of 3.5 GeV and a natural emittance of 3.9 nm·rad \cite{SSRF}.
Its storage ring consists of 4 SPs with 20 DBA cells.
Each SP has 3 standard cells and 2 matching cells.
The linear optical functions and magnet layout of a half SP are shown in Fig. \ref{fig:SSRF_optics}.
There are 2 chromatic sextupole families (SD and SF) in the high dispersion regions, and 6 harmonic sextupole families (S1-S6) in the relatively low dispersion regions.
The families S1, S3 and S5 are horizontally focusing sextupoles, and S2, S4 and S6 are defocusing ones.

In our nonlinear optimization, the strengths of six harmonic sextupole families are variables, with two chromatic sextupole families for fitting the corrected chromaticities to $(1, 1)$.
To statistically analyze the correlation between the RDTs and DA area, a large number of nonlinear solutions need to be generated.
The probability of finding a nonlinear solution with a large DA in a randomly generated solution set is very small. 
Now that minimizing the RDTs of one-turn map is a necessary condition for enlarging the DA, we can increase the proportion of nonlinear solutions with large DAs by minimizing the RDTs. 
With a genetic algorithm toolbox geatpy \cite{geatpy}, 10000 nonlinear solutions were obtained after 40 generations of minimizing the third-order RDTs, and the third-order RDTs of some solutions are almost completely cancelled.
Then the on-momentum DA areas of all nonlinear solutions were calculated with ELEGANT \cite{ELEGANT}.

\begin{figure}
    \includegraphics[width=\linewidth]{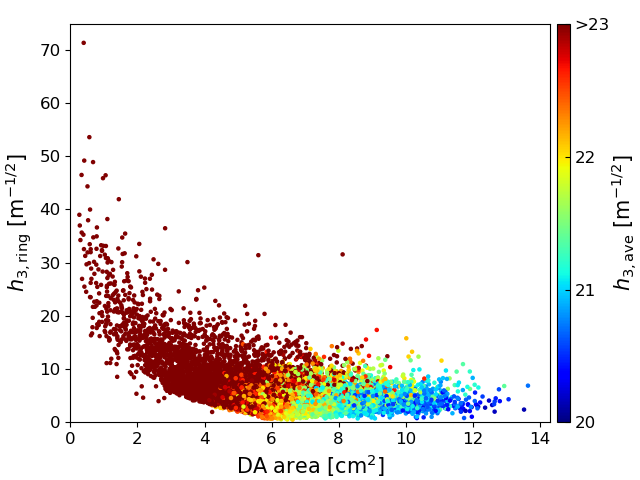}
    \caption{\label{fig:h3ring_area}Correlation between the DA area, the third-order RDTs $h_{3,\mathrm{ring}}$ and their fluctuations $h_{3,\mathrm{ave}}$ for the nonlinear solutions of the SSRF DBA lattice. Red color indicates large RDT fluctuations, and blue color indicates small RDT fluctuations.}
\end{figure}
\begin{figure*}
    \includegraphics[width=0.45\linewidth]{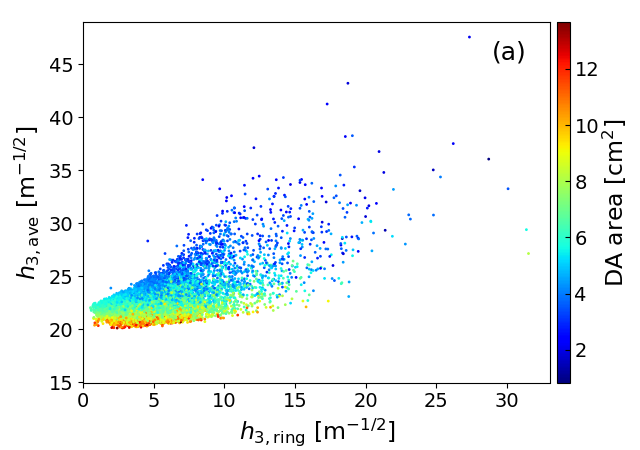}    \includegraphics[width=0.45\linewidth]{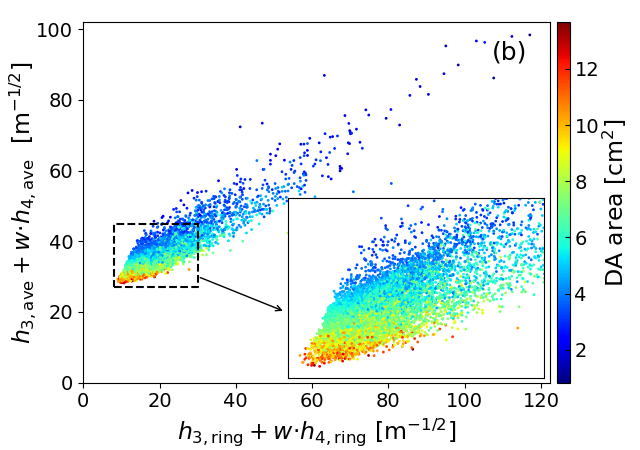}
    \includegraphics[width=0.45\linewidth]{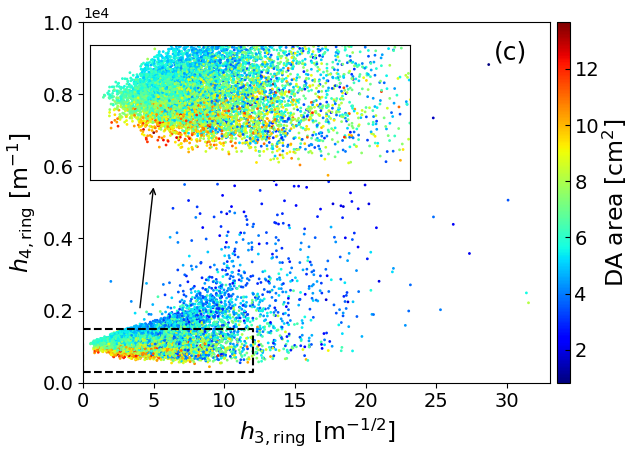}
    \includegraphics[width=0.45\linewidth]{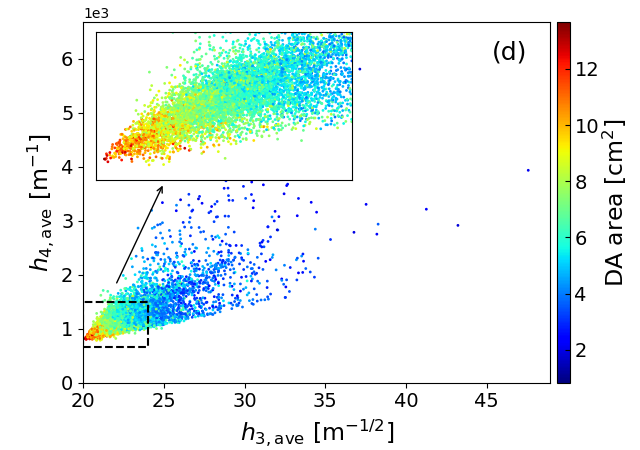}
    \caption{\label{fig:SSRF_comparison}Different correlations between the DA area, the third- and fourth-order RDTs of one-turn map and the fluctuations of these RDTs for the SSRF DBA lattice. Red color indicates large DA areas, and blue for small DA areas. The nonlinear solutions shown here have $h_{\mathrm{ADTS}} < 10000 \ \mathrm{m}^{-1}$.}
\end{figure*}
Following Ref. \cite{Yang2011}, we show the correlation between the DA area and the third-order RDTs of one-turn map, i.e. $h_{3,\mathrm{ring}}$, for these solutions in Fig. \ref{fig:h3ring_area}.
Besides, their RDT fluctuations $h_{3, \mathrm{ave}}$ are shown in the figure as a color bar.
The correlation between DA area and $h_{3,\mathrm{ring}}$ roughly follows what was found in Ref. \cite{Yang2011}: small $h_{3,\mathrm{ring}}$ is a necessary but not sufficient condition for large DA.
However, the DA area has a stronger correlation with $h_{3, \mathrm{ave}}$.
For a solution with small $h_{3, \mathrm{ave}}$, the probability of having a large DA area is larger than the solution with small $h_{3, \mathrm{ring}}$.
Therefore, minimizing the RDT fluctuations is more effective than minimizing RDTs themselves in enlarging the DA.
Besides, there is an interesting thing that for the solutions with small $h_{3, \mathrm{ave}}$, their $h_{3, \mathrm{ring}}$ are not large.

The third-order RDTs are the most important in this DBA lattice \cite{SSRF_Tian}.
For a more comprehensive comparison and a better understanding, the ADTS terms as well as the fourth-order RDTs were further involved in the nonlinear analysis.
Of the generated solutions, the solutions with $h_{\mathrm{ADTS}} < 10000 \ \mathrm{m}^{-1}$ were used for the further analysis.
For these solutions, Fig. \ref{fig:SSRF_comparison} shows different correlations between the DA area, the one-turn RDTs and the fluctuations of RDTs.
Figure \ref{fig:SSRF_comparison}(a) is another representation of Fig. \ref{fig:h3ring_area}, with the two axes representing $h_{\mathrm{3, ring}}$ and $h_{\mathrm{3, ave}}$ and the color bar representing DA area.
It can be clearly seen that the colors are roughly layered horizontally, with solutions having large DAs, indicated by the red color, sinking to the bottom.
In Fig. \ref{fig:SSRF_comparison}(b), the fourth-order RDTs are further involved with the weight coefficient $w=0.01 \ \mathrm{m}^{1/2}$.
The RDT fluctuations are still pronounced, with colors again roughly layered.
In the two lower plots, we step further to analyze the one-turn RDTs and the RDT fluctuations separately.
Figure \ref{fig:SSRF_comparison}(c) shows the correlation between the third- and fourth-order one-turn RDTs and DA area.
We can see that the colors are layered clearly only when $h_{3,\mathrm{ring}}$ is quite small.
But when $h_{3,\mathrm{ring}}$ is relatively larger, many solutions with large differences in DA area are mixed together.
This is because the third-order RDTs dominate in this DBA lattice, and the significance of $h_{4,\mathrm{ring}}$ emerges when $h_{3,\mathrm{ring}}$ is small.
Comparing Figs. \ref{fig:SSRF_comparison}(a) and \ref{fig:SSRF_comparison}(c), we can find that minimizing $h_{3, \mathrm{ave}}$ is even more effective than minimizing the fourth-order term $h_{4, \mathrm{ring}}$.
A possible explanation is that the crossing terms of lower-order RDTs generate higher-order RDTs, thus indicating a underlying connection between the higher-order RDTs and the fluctuation of lower-order RDTs.
We will further demonstrate it in the next paragraph.
In Fig. \ref{fig:SSRF_comparison}(d), the two axes are changed to $h_{3, \mathrm{ave}}$ and $h_{4, \mathrm{ave}}$.
We can see that from the upper right to the lower left, the DA areas of these solutions gradually increase, and that the solutions with large DAs are on the tip of the lower left corner.
This reflects that there is a strong positive correlation between the third-order and fourth-order RDT fluctuations in this lattice.

\begin{figure}
    \includegraphics[width=\linewidth]{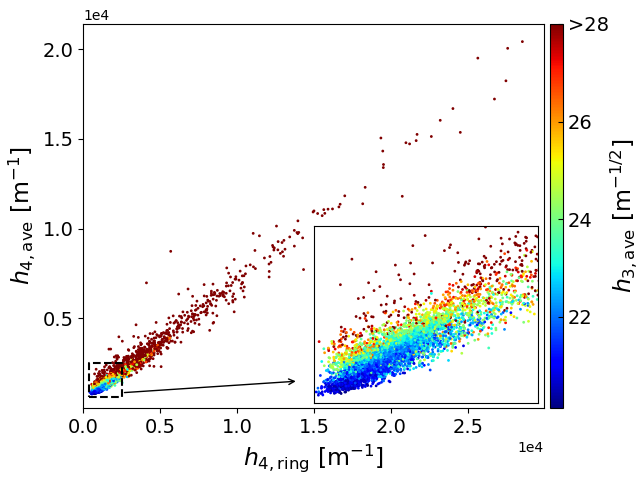}
    \caption{\label{fig:xxx}Correlation between $h_{3,\mathrm{ave}}$, $h_{4,\mathrm{ring}}$ and $h_{4, \mathrm{ave}}$ for the SSRF DBA lattice. Red color indicates large $h_{3, \mathrm{ave}}$, and blue color for small $h_{3,\mathrm{ave}}$.}
\end{figure}

Figure \ref{fig:xxx} shows the correlation between the higher-order $h_{4,\mathrm{ring}}$, $h_{4, \mathrm{ave}}$ and the lower-order $h_{3,\mathrm{ave}}$.
It is clear that both $h_{4,\mathrm{ring}}$ and $h_{4, \mathrm{ave}}$ roughly reduce as $h_{3,\mathrm{ave}}$ reduces.
This verifies that controlling the fluctuation of the third-order RDTs is beneficial for controlling the fourth-order RDTs and their fluctuations due to the cross-talk effect.
This is also consistent with Fig. \ref{fig:SSRF_comparison}(d).
Furthermore, the cross-terms can generate even higher-order RDTs, such as fifth-order RDTs, which are not only numerous in quantity, but also complicated to compute.
It is cumbersome to directly reduce them.
Therefore, we can minimize the fluctuations of third- and fourth-order RDTs to control them.

\begin{figure*}
    \includegraphics[width=0.43\textwidth]{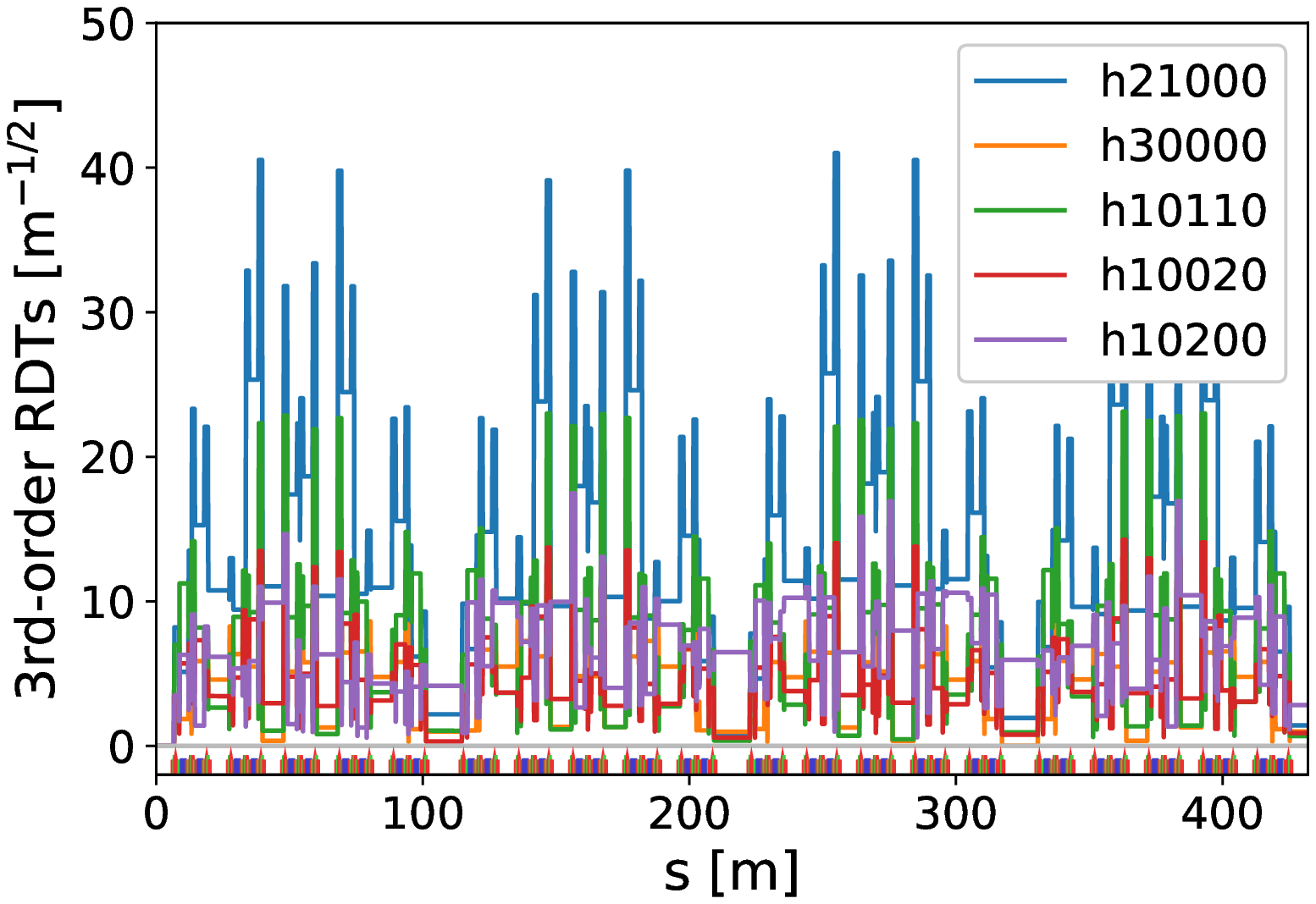}
    \includegraphics[width=0.43\textwidth]{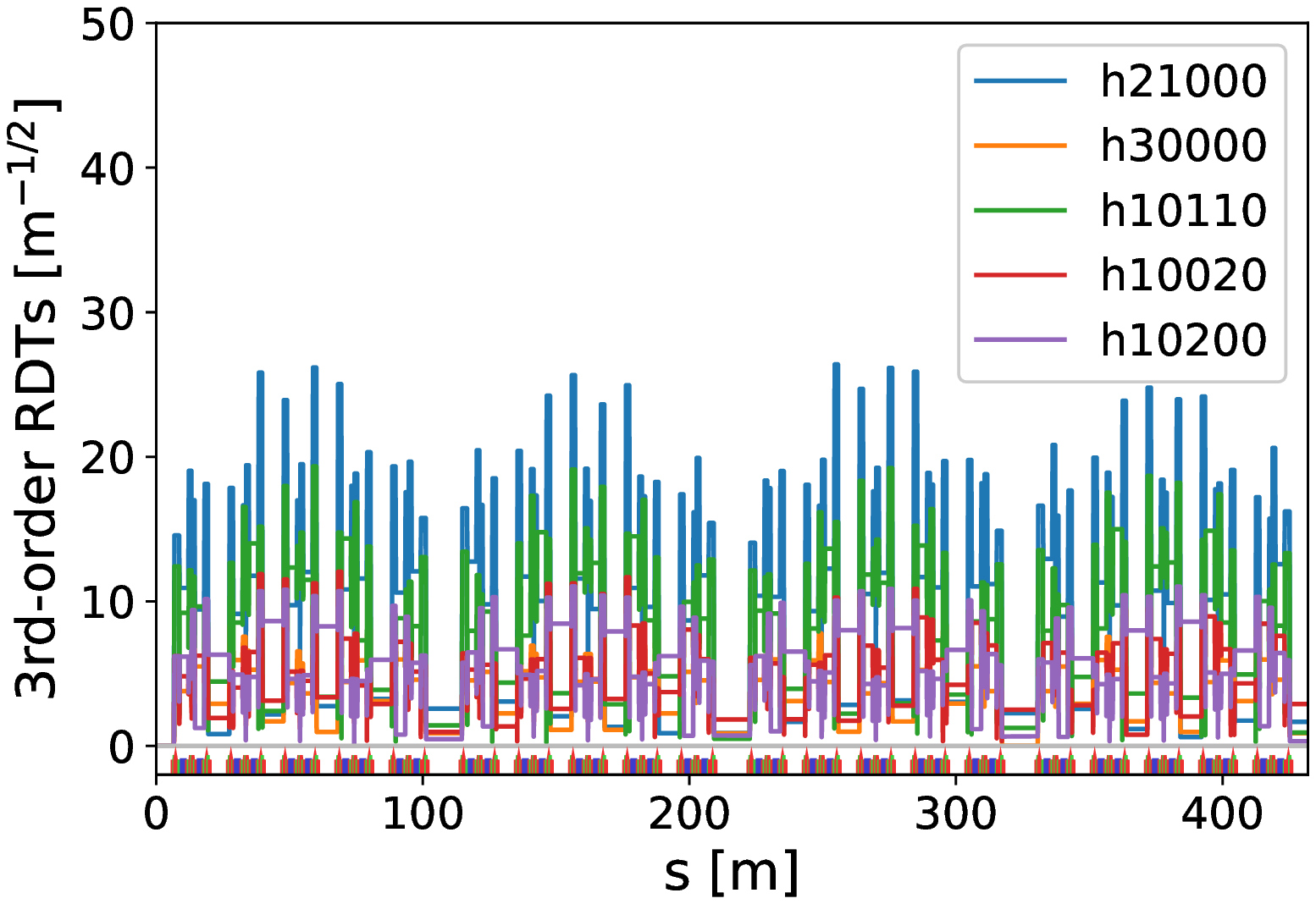}
    \includegraphics[width=0.43\textwidth]{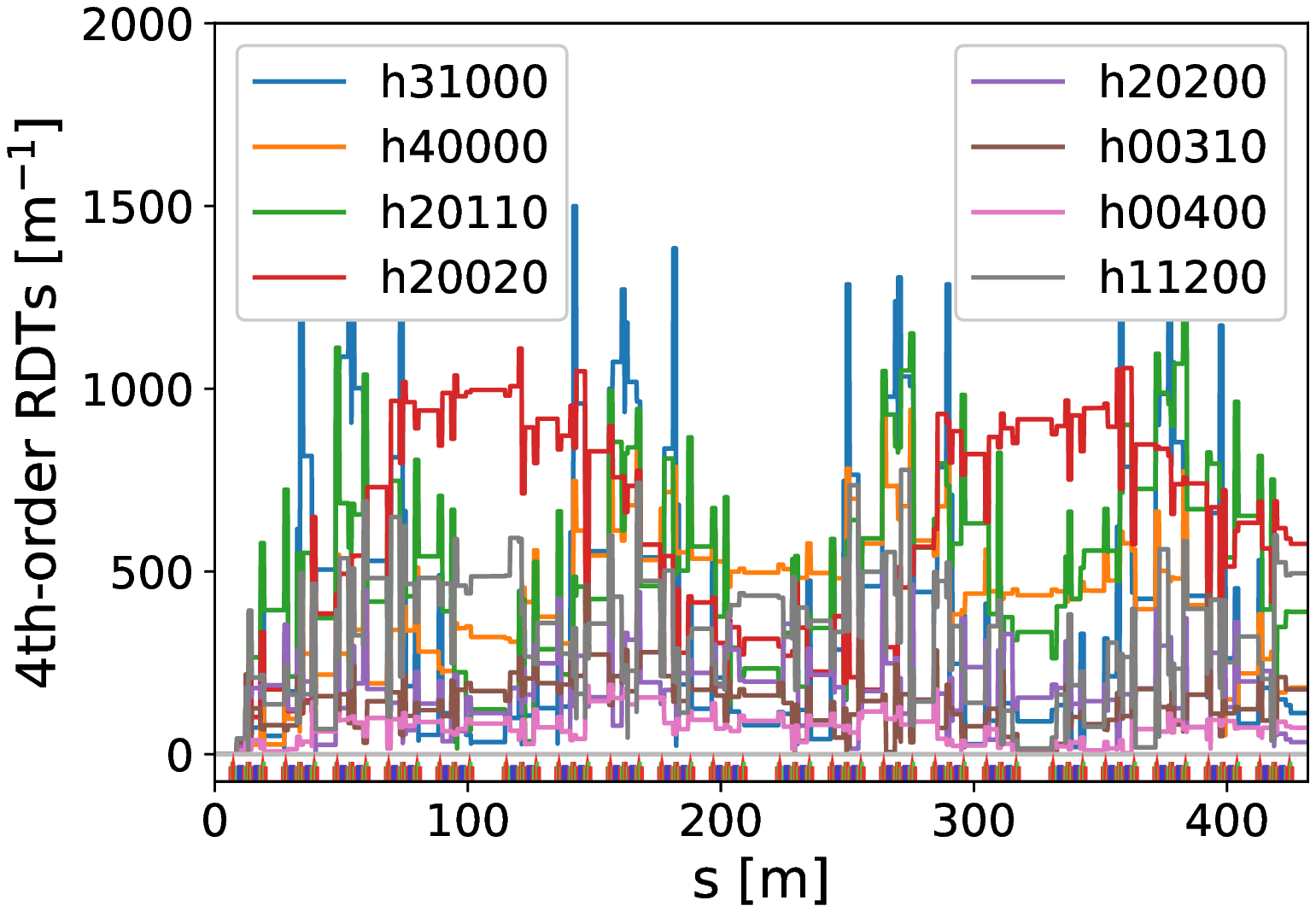}
    \includegraphics[width=0.43\textwidth]{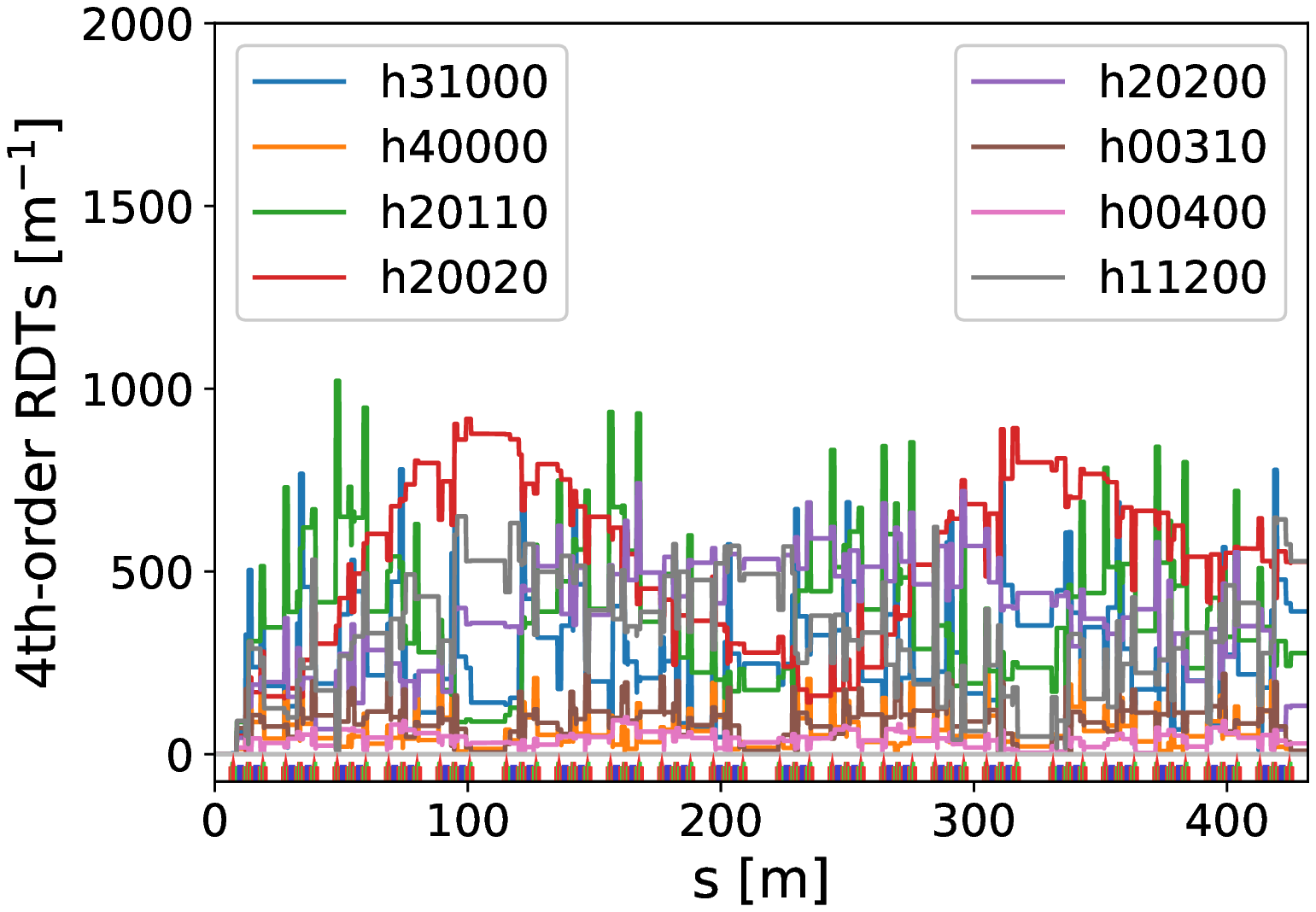}
    \includegraphics[width=0.45\textwidth]{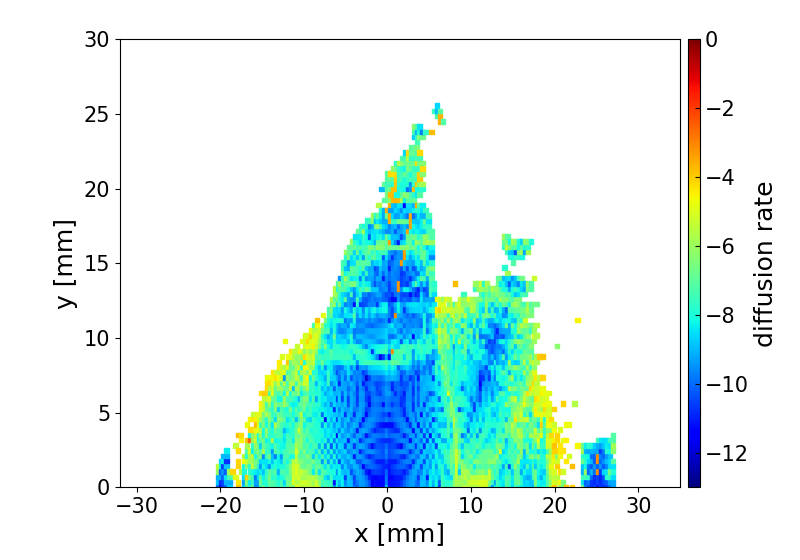}
    \includegraphics[width=0.45\textwidth]{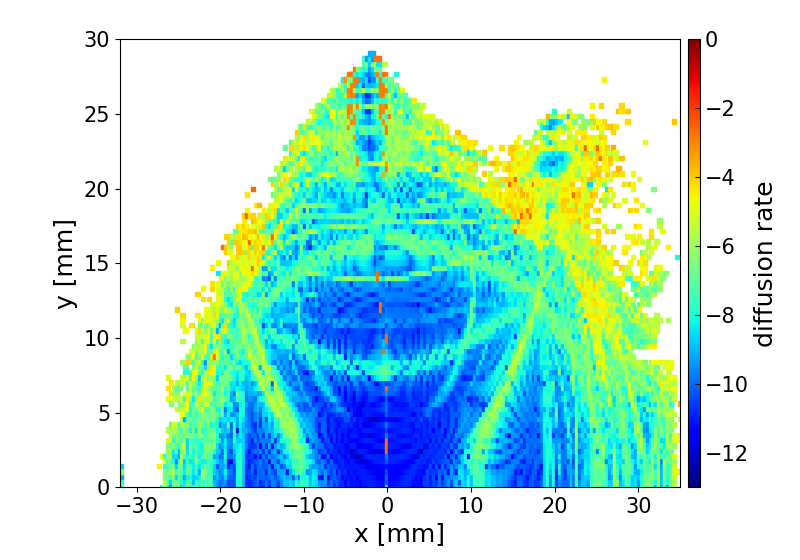}
    \includegraphics[width=0.45\textwidth]{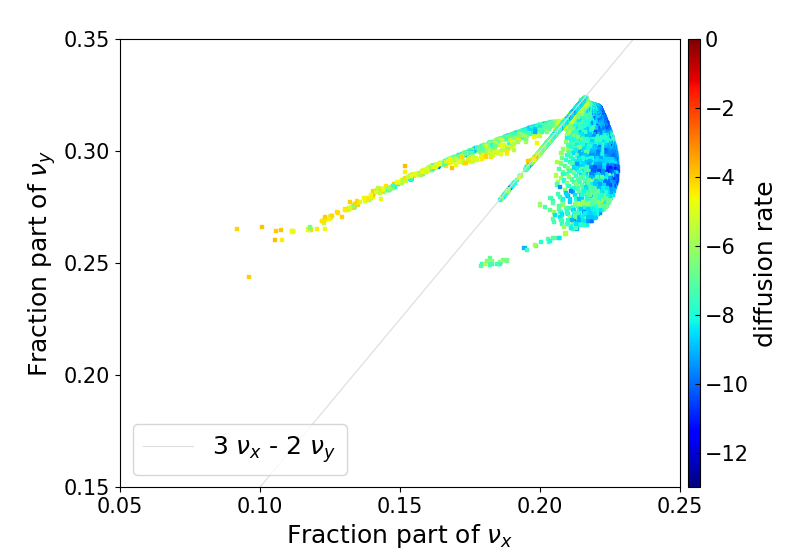}
    \includegraphics[width=0.45\textwidth]{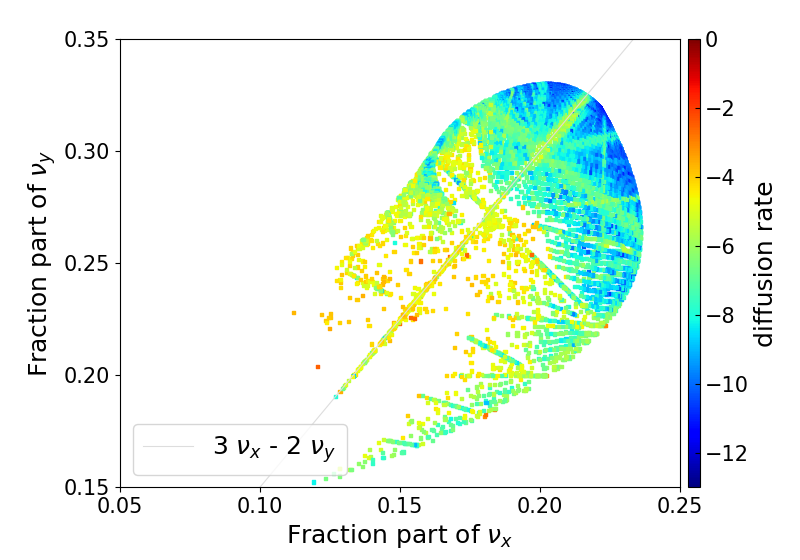}
    \caption{\label{fig:SSRF_examples}Two nonlinear solutions of the SSRF DBA lattice with similar one-turn RDT and ADTS terms but differnt RDT fluctuations and DA areas. The left four plots are for one solution, and the right four for the other solution. The upper four plots are the third- and fourth-order RDTs and their fluctuations, and the lower four plots are the frequency map analysis of DAs. The values of these nonlinear terms are listed in Table \ref{tab:SSRF_examples}.}
\end{figure*}

We pick out two nonlinear solutions with approximately the same values of $h_{3,\mathrm{ring}}$, $h_{4,\mathrm{ring}}$ and ADTS terms, but their $h_{3,\mathrm{ave}}$ and $h_{4,\mathrm{ave}}$ are different.
Table \ref{tab:SSRF_examples} shows these values of the two solutions.
Their DAs with frequency map analysis \cite{FMA} are tracked with ELEGANT and shown in Fig. \ref{fig:SSRF_examples}. The one-turn RDTs and the RDT fluctuations are also shown in the figure.
\begin{table} 
    \caption{\label{tab:SSRF_examples}%
    Nonlinear term values of two nonlinear solutions with similar RDT $h_{3,\mathrm{ring}}$, $h_{4,\mathrm{ring}}$ and ADTS terms but different RDT fluctuations $h_{3,\mathrm{ave}}$ and $h_{4,\mathrm{ave}}$.
    }
    \begin{ruledtabular}
    \begin{tabular}{ccc}
    &
    \textrm{smaller DA}&
    \textrm{larger DA}\\
    \colrule
    $h_{3,\mathrm{ring}}$  [m$^{-\frac{1}{2}}$]   & 3.5     & 3.6     \\
    $h_{4,\mathrm{ring}}\ [\textrm{m}^{-1}]$   & 901   & 897   \\
    $d\nu_x/dJ_x$ [m$^{-1}$] & -315 & -737 \\
    $d\nu_x/dJ_y$ [m$^{-1}$] & 857   & 1171   \\
    $d\nu_y/dJ_y$ [m$^{-1}$] & -2082 & -2472 \\
    $h_{3,\mathrm{ave}}$  [m$^{-\frac{1}{2}}$] & 23.3    & 20.7    \\
    $h_{4,\mathrm{ave}}$ [m$^{-1}$]  & 1137  & 916   \\
    \end{tabular}
    \end{ruledtabular}
\end{table}
We can clearly see that the solution with smaller RDT fluctuations, i.e. smaller $h_{3,\mathrm{ave}}$ and $h_{4,\mathrm{ave}}$, has a larger DA.
For the solution with smaller DA, the fifth-order resonance line $3 \nu_x - 2 \nu_y$ has a more significant effect.
This verifies that controlling the fluctuations of third- and fourth-order RDTs is beneficial for controlling the fifth-order RDTs.

\section{\label{sec:IV}Optimization of 6BA lattices}
To achieve a diffraction-limited emittance with a reasonable circumference, MBA lattices are used in the design of DLSRs to replace DBA lattices \cite{Hettel2014, Einfeld2014}.
In this section, we will use DLSR MBA lattices to study the correlation between RDT fluctuations and DA area again.
In these lattices with strong focusing, the nonlinear effects become strong.
HOA is a successful approach to control the nonlinear effects and has been used in some DLSR lattice designs \cite{sls2, BengtssonIPAC2019, SOLEIL, Yang2021, renipac2021, SKIF}.
In an HOA MBA lattice with appropriate bend unit cell tunes, most or all of the third- and fourth-order geometric RDTs can be cancelled over some identical cells \cite{Bengtsson2017}.
Here the MBA lattices used are two HOA 6BA lattices that we designed in Refs. \cite{1st6BA, Yang2021}, which have five identical unit cells, each with horizontal and vertical tunes of $(0.4, 0.1)$.
In this kind of HOA lattices, the fourth-order RDT $h_{20200}$ can not be cancelled in the ideal cancellation condition \cite{Bengtsson2017}.
Besides, in these 6BA lattices, the HOA approach was also used for further nonlinear cancellation over some lattice cells.
For one of the 6BA lattices, the term $h_{20200}$ is still not cancelled over some lattice cells; while for the other one, $h_{20200}$ is cancelled over lattice cells.

\subsection{\label{sec:IVa}The first 6BA lattice}
\begin{figure}
    \includegraphics[width=\linewidth]{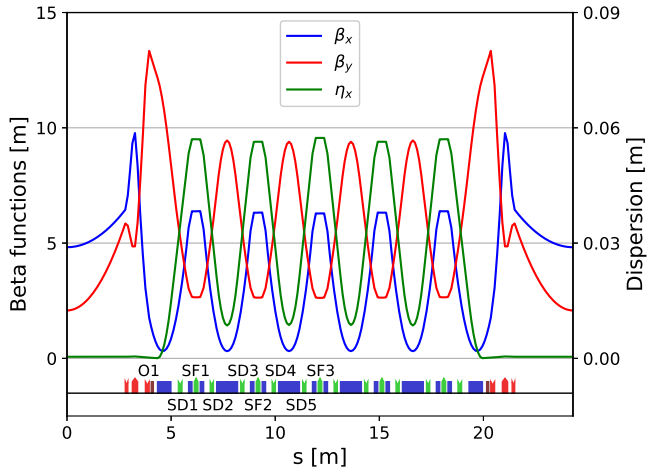}
    \caption{\label{fig:HOA6BA16cells}Linear optical functions and magnet layout of one cell of the first 6BA lattice.
    In the layout, bends are in blue, quadrupoles in red, sextupoles in green and octupoles in brown.}
\end{figure}

The first 6BA lattice we will study was designed in Ref. \cite{1st6BA}. The designed storage ring is a 2.2 GeV DLSR with a natural emittance of 94 pm·rad, which consists of 16 identical lattice cells.
The optical functions of this lattice are shown in Fig. \ref{fig:HOA6BA16cells}.
The horizontal and vertical tunes of a lattice cell are close to $(2+5/8, 7/8)$, enabling nonlinear cancellation over 8 cells.
However, neither the lattice cell tunes nor the unit cell tunes are able to cancel $h_{20200}$.
To further optimize the nonlinear dynamics, including the control of ADTS terms, the sextupoles are symmetrically grouped into 8 families as illustrated in Fig. \ref{fig:HOA6BA16cells}, and a family of octupoles is used as in Ref. \cite{1st6BA}. 

\begin{figure}
    \includegraphics[width=\linewidth]{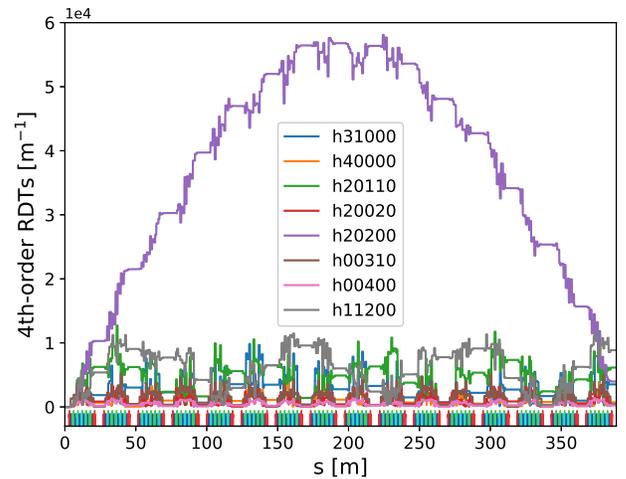}
    \caption{\label{fig:16cellsExample}The fourth-order RDT fluctuations along the first 6BA ring of a nonlinear solution. 
    The RDT $h_{20200}$ rises very high.}
\end{figure}

Similar to the DBA lattice, we use genetic algorithm to increase the proportion of solutions with good dynamic performance for better nonlinear analysis.
But here three objectives $h_{3,\mathrm{ring}}$, $h_{4,\mathrm{ring}}$ and $h_\mathrm{ADTS}$ were optimized simultaneously, since fourth-order RDTs and ADTS terms become more important in the nonlinear optimization of DLSR lattices.
The chromaticities were corrected to $(2, 2)$.
The genetic algorithm ran 10 generations with a population size of 10000.
The fourth-order RDT fluctuations of one optimized solution along the ring are shown in Fig. \ref{fig:16cellsExample}.
The term $h_{20200}$ exhibits a significant rise in magnitude along the ring, while the other terms are well suppressed.
As previously mentioned, the term $h_{20200}$ cannot be cancelled in two HOA schemes, i.e. nonlinear cancellation within a single lattice cell and over 8 lattice cells, and can only be controlled through the nonlinear optimization with sextupole grouping.

\begin{figure}
    \includegraphics[width=0.9\linewidth]{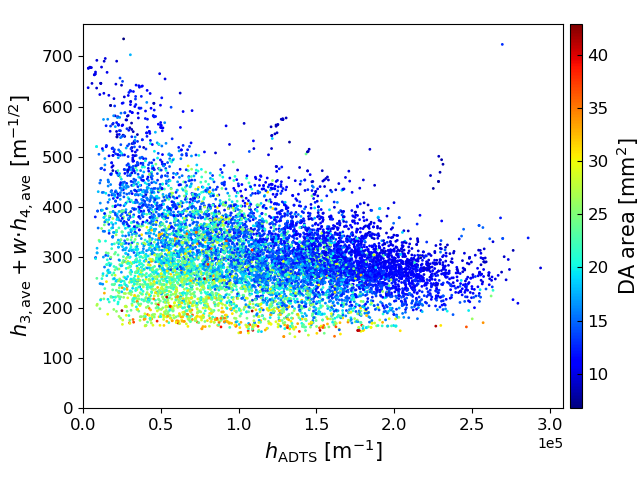}
    \caption{\label{fig:1st6BA_adts}Correlation between the ADTS terms, the RDT fluctuations and the DA areas for the first 6BA lattice.}
\end{figure}

\begin{figure}
    \includegraphics[width=\linewidth]{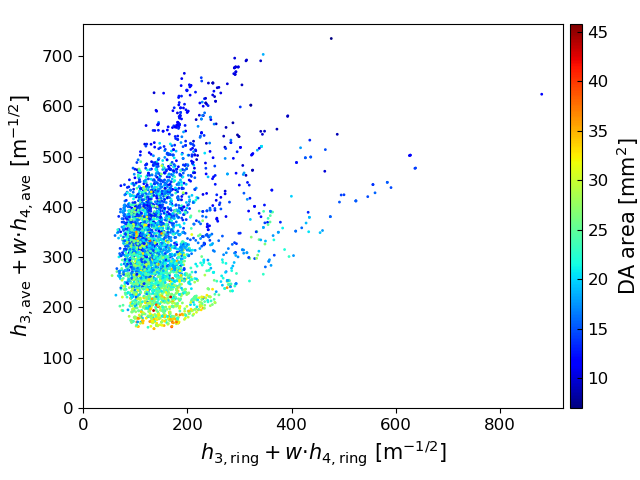}
    \includegraphics[width=\linewidth]{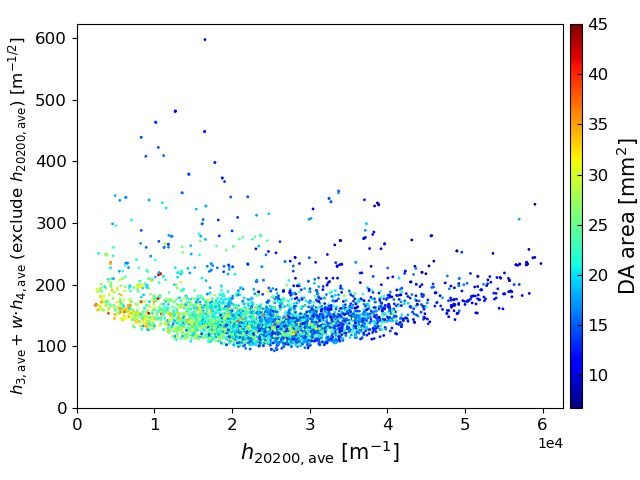}
    \caption{\label{fig:HOA6BA16cellsComparison}Upper plot: correlation between the DA area, the one-turn RDTs and the RDT fluctuations for the first 6BA.
    Lower plot: correlation between the DA area, the fluctuation of $h_{20200}$ and other RDT fluctuations.
    The nonlinear solutions shown here have $h_{\mathrm{ADTS}} < 1 \times 10^5 \ \mathrm{m}^{-1}$.}
\end{figure}

For the optimized solutions, their $h_{3, \mathrm{ave}}$, $h_{4,\mathrm{ave}}$ and DA areas were calculated.
Figure \ref{fig:1st6BA_adts} shows the correlation between the RDT fluctuations, ADTS terms and DA area.
The weight coefficient $w$ is also set to $0.01 \ \mathrm{m}^{-1}$ in this lattice.
Compared to the DBA lattice, ADTS terms are more difficult to control in this 6BA lattice with stronger focusing.
Most of the solutions with large ADTS values have small DA areas.
For the solutions with both small ADTS values and small RDT fluctuations, most of them have large DAs.
We use the solutions with $h_{\mathrm{ADTS}} < 1\times 10^{5} \ \mathrm{m}^{-1}$ for further analysis.
The correlation between RDT fluctuations and one-turn RDTs is shown in the upper plot of Fig. \ref{fig:HOA6BA16cellsComparison}.
We can see that similar to the DBA lattice case, DA area has a stronger correlation with RDT fluctuations than one-turn RDTs, with the red color sinking to the bottom.
Besides, the differences in RDT fluctuations of these solutions can be large when their one-turn RDTs are controlled.
The lower plot of Fig. \ref{fig:HOA6BA16cellsComparison} shows that the term $h_{20200,\mathrm{ave}}$ contributes the main difference.
We can see that for the solutions with small $h_{20200,\mathrm{ave}}$, most of them have large DA areas.
And for the solutions with small $h_{20200,\mathrm{ave}}$ but large fluctuations of other RDTs, their DAs are small, indicating that controlling the fluctuation of other RDTs is also important.

\subsection{The second 6BA lattice}
\begin{figure}
    \includegraphics[width=\linewidth]{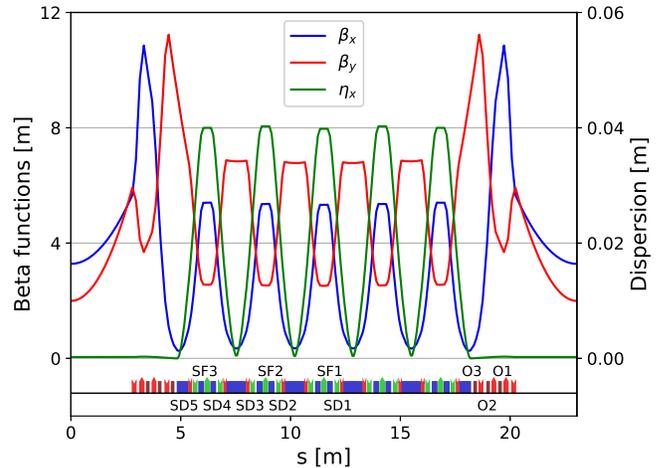}
    \caption{\label{fig:HOA6BA20cells}Linear optical functions and magnet layout of the second 6BA lattice cell.}
\end{figure}

The second 6BA lattice to be studied was designed in Ref. \cite{Yang2021}. The beam energy is also 2.2 GeV.
But the storage ring consists of 20 identical lattice cells, and has a lower natural emittance of 36 pm·rad and lower beta functions in straight sections.
Figure \ref{fig:HOA6BA20cells} shows one cell of this lattice.
Different from the first 6BA lattice, the horizontal and vertical tunes of this lattice cell are approximately $(2.7, 0.9)$ to make the nonlinear cancellation over 10 cells, including the cancellation of $h_{20200}$.
In this lattice, the sextupoles are also symmetrically grouped as shown in Fig. \ref{fig:HOA6BA20cells}.
There are 8 families of chromatic sextupoles and 3 families of harmonic octupoles used for the nonlinear optimization, with the chromaticities corrected to $(-3, -3)$ due to negative momentum compaction factor.

\begin{figure}
    \includegraphics[width=\linewidth]{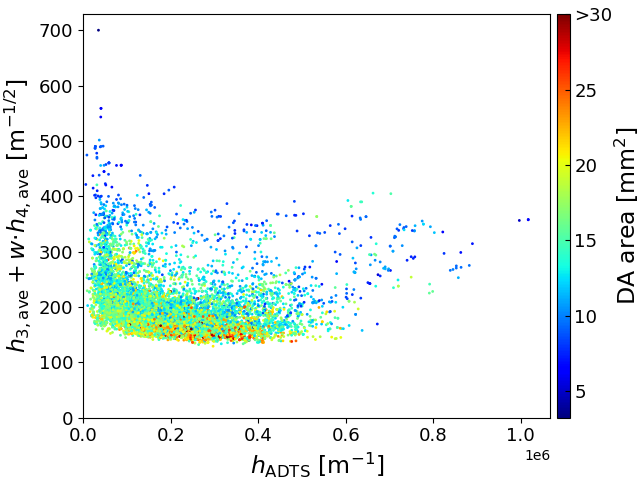}
    \caption{\label{fig:2nd6BA_adts}Correlation between the ADTS terms, the RDT fluctuations and the DA areas for the second 6BA lattice.}
\end{figure}

\begin{figure}
    \includegraphics[width=\linewidth]{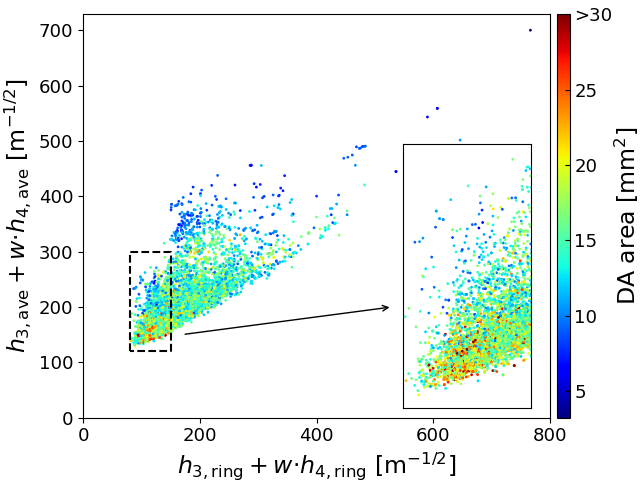}
    \includegraphics[width=\linewidth]{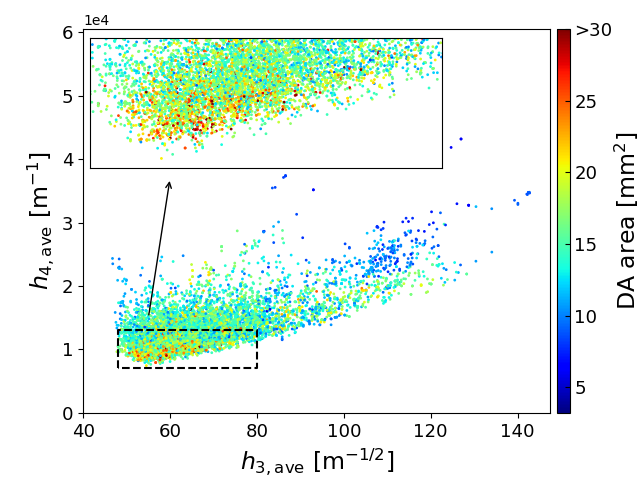}
    \caption{\label{fig:HOA6BA20cellsComparison}Upper plot: correlation between the DA area, the one-turn RDTs and the RDT fluctuations for the 2nd 6BA.
    Lower plot: correlation between the DA area, the third- and fourth-order RDT fluctuations with different orders being considered separately.
    The nonlinear solutions shown here have $h_{\mathrm{ADTS}} < 5 \times 10^{5} \ \mathrm{m}^{-1}$.}
\end{figure}

For the nonlinear analysis, the three objectives $h_{3,\mathrm{ring}}$, $h_{4,\mathrm{ring}}$ and $h_\mathrm{ADTS}$ were also optimized with a population of 10000 and 20 generations here.
The correlation between the RDT fluctuations, ADTS terms and DA area is shown in Fig. \ref{fig:2nd6BA_adts}.
The weight coefficient $w = 0.01 \ \mathrm{m}^{-1}$.
Compared to the first 6BA lattice, this lattice has stronger nonlinear effects with larger ADTS terms.
Nonetheless, possibly due to the effective suppression of resonances with the HOA strategy, even if the ADTS terms are large, there are still some solutions with large DAs.
Next we analyze the solutions with $h_\mathrm{ADTS} < 5 \times 10^{5} \ \mathrm{m}^{-1}$.
The upper plot of Fig. \ref{fig:HOA6BA20cellsComparison} shows the correlation between one-turn RDTs, RDT fluctuations and DA area.
Compared to the first 6BA, here the range of $h_{3, \mathrm{ave}} + w\cdot h_{4, \mathrm{ave}}$ of the solutions is smaller when one-turn RDTs are controlled.
This is because $h_{20200}$ is prevented from building up in this lattice.
And we can see that the solutions with large DAs are mainly at the bottom, indicating the effectiveness of minimizing RDT fluctuations.
In the lower plot, just like Fig. \ref{fig:SSRF_comparison}(d), we further analyze the third- and fourth-order RDT fluctuations separately.
Most solutions with large DAs have small $h_{3, \mathrm{ave}}$ and $h_{4, \mathrm{ave}}$, which is consistent with Fig. \ref{fig:SSRF_comparison}(d).
But different from Fig. \ref{fig:SSRF_comparison}(d), there are also some solutions with small $h_{3, \mathrm{ave}}$ and $h_{4, \mathrm{ave}}$ have small DAs.
This needs to be further studied.
We have preliminarily found that optimizing the weight of each RDT can strengthen the correlation between RDT fluctuations and DA area, since different resonances have different effects on DA.

\begin{figure}
    \includegraphics[width=\linewidth]{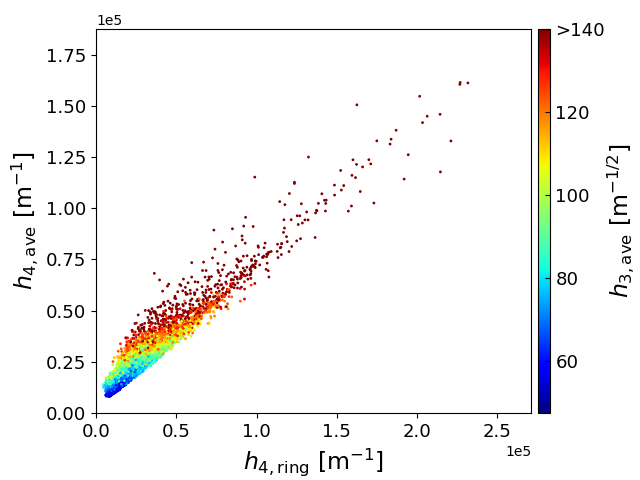}
    \caption{\label{fig:2nd6BA_xxx}Correlation between $h_{3,\mathrm{ave}}$, $h_{4,\mathrm{ring}}$ and $h_{4, \mathrm{ave}}$ for the second 6BA lattice.}
\end{figure}

We have shown that reducing the lower-order RDT fluctuations is beneficial for controlling higher-order RDTs in the DBA lattice.
But in this 6BA lattice, the fourth-order RDT fluctuations are contributed not only by the crossing terms of sextupoles, but also by the octupoles, and $h_{3, \mathrm{ave}}$ only affects the former.
To verify the correlation between third-order RDT fluctuations and fourth-order RDTs in this lattice like in Fig. \ref{fig:xxx}, we generated another set of nonlinear solutions by optimizing $h_{3, \mathrm{ave}}$ for some generations, where the octupoles were not employed.
For these solutions, the ones with smaller $h_{3, \mathrm{ave}}$ also have smaller $h_{4, \mathrm{ave}}$ and $h_{4, \mathrm{ring}}$, as shown in Fig. \ref{fig:2nd6BA_xxx}.
This is consistent with Fig. \ref{fig:xxx}.
In the lower plot of Fig. \ref{fig:HOA6BA20cellsComparison}, the correlation between $h_{3, \mathrm{ave}}$ and $h_{4, \mathrm{ave}}$ is weaker than that in Fig. \ref{fig:2nd6BA_xxx} due to that the octupoles also contribute to fourth-order RDTs.

\section{\label{sec:V}conclusion and outlook}
Inspired by the fact that the local cancellation of nonlinear dynamics effects adopted in some DLSR lattices is more effective than the global cancellation, we studied the analysis of nonlinear dynamics based on minimizing the RDT fluctuations along the ring.
A DBA lattice and two 6BA lattices were taken as examples for this study.
We calculated the RDTs as functions of position along the ring, and the RDT fluctuations were characterized by using the average RDT values at nonlinear magnet locations.
It was found that reducing the RDT fluctuations has a very strong correlation with enlarging the DA area.
Nonlinear solutions with small RDT fluctuations are much more likely to have large DAs than those with small one-turn RDTs.
And for the solutions with small RDT fluctuations, their one-turn RDTs are also controlled.
Moreover, reducing the fluctuation of lower-order RDTs can also reduce the higher-order RDTs and their fluctuations.
The higher-order RDTs contributed by the crossing terms of lower-order RDTs are not only numerous but also computationally complicated, especially for the fifth-order and higher-order RDTs.
The fifth-order case was demonstrated in the DBA lattice.
The effectiveness of controlling RDT fluctuations in enlarging DA confirms once again that the local nonlinear cancellation is more effective than the global cancellation.

Since reducing the RDT fluctuations can enlarge the DA area more effectively than reducing the one-turn RDTs, we can consider minimizing RDT fluctuations in the DA optimization.
By using evolutionary algorithms, we can first minimize RDT fluctuations to effectively and quickly find the regions where large DA solutions exist, and then in these regions, DA can be further optimized based on particle tracking.
Although this paper focused on on-momentum DA and the fluctuation of geometric RDTs, it is possible that the fluctuation of chromatic terms related to off-momentum dynamics can be further considered in the nonlinear optimization.
In addition, the RDT fluctuations can provide physical feedback for adjusting linear optics to achieve better nonlinear dynamics performance.

The quality of DA is related to both the area of the DA and the diffusion rate inside the DA.
Lower diffusion rates indicate that the motion of particles is more regular \cite{FMA} and the DA has better robustness against errors.
Since reducing RDT fluctuations can control both lower-order and higher-order resonances, it may lead to lower diffusion rates.
Therefore, we will further study the correlation between RDT fluctuations, DA area and diffusion rates using frequency map analysis.
Besides, machine learning, which has been successfully applied to the nonlinear dynamics optimization in recent years \cite{LiYJ2018ML,WAN2019,Wan2020ML,APS2021,CERN_ML,Wan_2022}, can also be used to enhance the study in this paper, including better characterization of the RDT fluctuations.
Since reducing RDT fluctuations is more effective than reducing one-turn or one-period RDTs, we can explore new lattices based on minimizing the RDT fluctuations in the linear and nonlinear optimization of a general magnet layout.

\begin{acknowledgments}
One of the authors (Zhenghe Bai) would like to thank Laurent Nadolski and Ryutaro Nagaoka of SOLEIL for helpful discussions.
This work was supported by the National Natural Science Foundation of China under Grant No. 11875259 and the National Key Research and Development Program of China under Grant No. 2016YFA0402000.
\end{acknowledgments}

\appendix*
\section{\label{appendix}Fluctuation of the fourth-order RDTs}
As in Sec.~\ref{sec:II}, here we also use $\bm{m} = (j - k, l - m)$ to represent the mode of $h_{jklm0}$, which drives the resonance $(j - k) \nu_x + (l - m)\nu_y$.
For simplicity, we substitute $h_{jklm0}$ with $h_{\bm{m}}$.
The fourth-order RDTs are contributed by octupoles and the crossing terms of sextupoles.
The fluctuation of the fourth-order RDTs contributed by octupoles is as simple as the third-order RDTs in Sec.~\ref{sec:II} \cite{Octupole}, and with the number of lattice periods $u$ as a variable, the fluctuation of such a RDT can be described by a constant term and the $e^{iu\bm{m}\cdot \bm{\mu}}$ term.
In the following we will characterize the fluctuation of the fourth-order RDTs contributed by the crossing terms of sextupoles.

For a lattice period with $N$ sextupoles, 
we denote the period tunes as $\bm{\mu} = 2 \pi (\nu_x, \nu_y)$ and the phase advances as $\bm{\phi} = (\phi_x, \phi_y)$. 
And we use $h_{t, \bm{m}}$ to represent the contribution of the $t$-th sextupole to the third-order RDT.
The sextupole terms $h_{a, \bm{m}_1}$ and $h_{b, \bm{m}_2}$  drive the fourth-order resonance $\bm{m}_1 + \bm{m}_2$ by the cross-talk effect \cite{SLS97}:
\begin{eqnarray}\label{crossing}
    \frac{1}{2}\sum_{b>a=1}^{N}\left[h_{a, \bm{m}_1}(2J_x)^{\frac{j_1 + k_1}{2}}(2J_y)^{\frac{l_1 +m_1}{2}}e^{i\bm{m}_1\bm{\phi}}, \right.\nonumber\\
    \left.h_{b, \bm{m}_2}(2J_x)^{\frac{j_2 + k_2}{2}}(2J_y)^{\frac{l_2 +m_2}{2}}e^{i\bm{m}_2\bm{\phi}}\right].
\end{eqnarray}
We move the terms $h_{a, \bm{m}_1}$ and $h_{b, \bm{m}_2}$ outside the Poisson bracket, then
the coefficient of the Poisson bracket is $\frac{1}{2}\sum_{b>a=1}^{N} h_{a, \bm{m}_1}h_{b, \bm{m}_2}$.

The crossing term fluctuation arises from the change of $\sum_{b>a=1}^{t} h_{a, \bm{m}_1}h_{b, \bm{m}_2}$ with $t$.
For multiple lattice periods, at the $(u+1)$-th period $(u\geq 0)$, 
\begin{eqnarray} \label{multi-period_4th}
    &\sum_{b>a=1}^{u\cdot N + t} h_{a, \bm{m}_1}h_{b, \bm{m}_2} 
    = \sum_{b=2}^{u\cdot N + t}\sum_{a=1}^{b-1}h_{a, \bm{m}_1}h_{b, \bm{m}_2}\nonumber \\ 
        \nonumber \\
    = & \sum_{b=2}^{u\cdot N}\sum_{a=1}^{b-1}h_{a, \bm{m}_1}h_{b, \bm{m}_2} 
    + \sum_{b=u\cdot N + 1}^{u\cdot N + t}\sum_{a=1}^{b-1}h_{a, \bm{m}_1}h_{b, \bm{m}_2}
    \nonumber \\ 
\end{eqnarray}
We denote that $h_{a,\bm{m}_1}$ is at the $(u_{a}+1)$-th period and $h_{b,\bm{m}_2}$ is at the $(u_b+1)$-th period, with $u_b \geq u_a \geq 0$.
The first part in Eq.~(\ref{multi-period_4th}) can be divided into two parts with $u_b = u_a$ and $u_b > u_a$.
Then we have
\begin{widetext}
    \begin{eqnarray} \label{eq.1st}
        \sum_{b=2}^{u\cdot N}\sum_{a=1}^{b-1}h_{a, \bm{m}_1}h_{b, \bm{m}_2} =& \sum_{u_{a} =0}^{u-1}\sum_{b>a=1}^{N} h_{(u_a \cdot N + a), \bm{m}_1} h_{(u_a \cdot N + b),\bm{m}_2} + \sum_{u_b>u_a=0}^{u-1}\sum_{a,b=1}^{N} h_{(u_a \cdot N + a), \bm{m}_1}h_{(u_b \cdot N + b), \bm{m}_2} \nonumber \\
        = &\sum_{u_{a} =0}^{u-1}\sum_{b>a=1}^{N} h_{a, \bm{m}_1} h_{b,\bm{m}_2} e^{iu_{a} (\bm{m}_1 + \bm{m}_2)\cdot \bm{\mu}} + \sum_{u_b>u_a=0}^{u-1}\sum_{a,b=1}^{N} h_{a, \bm{m}_1}h_{b, \bm{m}_2}e^{i(u_a \bm{m}_1 + u_b \bm{m}_2)\cdot \bm{\mu}} \nonumber \\
        =&\left(\sum_{b>a=1}^{N} h_{a, \bm{m}_1}h_{b, \bm{m}_2}\right) \frac{1 - e^{iu(\bm{m}_1 + \bm{m}_2)\cdot \bm{\mu}}}{1 - e^{i(\bm{m}_1 + \bm{m}_2)\cdot \bm{\mu}}} + \left(\sum_{a,b=1}^{N}h_{a, \bm{m}_1}h_{b, \bm{m}_2}\right)\sum_{u_b=1}^{u-1}e^{i u_{b} \bm{m}_2\cdot \bm{\mu}}\frac{1 - e^{iu_{b}\bm{m}_1\cdot \bm{\mu}}}{1 - e^{i\bm{m}_1\cdot \bm{\mu}}} \nonumber \\
        =&\left(\sum_{b>a=1}^{N} h_{a, \bm{m}_1}h_{b, \bm{m}_2}\right) \frac{1 - e^{iu (\bm{m}_1 + \bm{m}_2)\cdot \bm{\mu}}}{1 - e^{i(\bm{m}_1 + \bm{m}_2)\cdot \bm{\mu}}} + \frac{\sum_{a,b=1}^{N}h_{a, \bm{m}_1}h_{b, \bm{m}_2}}{1 - e^{i\bm{m}_1\cdot \bm{\mu}}}\left(
        \frac{1 - e^{i u \bm{m}_{2}\cdot \bm{\mu}}}{1 - e^{i\bm{m}_{2}\cdot \bm{\mu}}} - \frac{1 - e^{i u (\bm{m}_{1} + \bm{m}_{2})\cdot \bm{\mu}}}{1 - e^{i(\bm{m}_{1} + \bm{m}_{2})\cdot \bm{\mu}}}\right),
        \nonumber \\
    \end{eqnarray}
\end{widetext}
where $h_{(u \cdot N + t), \bm{m}} = h_{t, \bm{m}} e^{i u \bm{m}\cdot \bm{\mu}}$ derived in Ref.~\cite{CERN8805}.
And the second part in Eq.~(\ref{multi-period_4th}) is
\begin{widetext}
    \begin{eqnarray} \label{eq.2nd}
        \sum_{b=u\cdot N + 1}^{u\cdot N + t}\sum_{a=1}^{b-1}h_{a, \bm{m}_1}h_{b, \bm{m}_2} 
        =&\sum_{b=u\cdot N+1}^{u\cdot N + t}\sum_{a=1}^{u\cdot N}h_{a, \bm{m}_1}h_{b, \bm{m}_2} + \sum_{b=u\cdot N+2}^{u\cdot N + t}\sum_{a=u\cdot N + 1}^{b-1}h_{a, \bm{m}_1}h_{b, \bm{m}_2} \nonumber \\
         =&\left(\sum_{a=1}^{N}h_{a, \bm{m}_1} \frac{1 - e^{iu \bm{m}_1\cdot \bm{\mu}}}{1 - e^{i\bm{m}_1\cdot \bm{\mu}}}\right)\left(\sum_{b=1}^{t}h_{b, \bm{m}_2}e^{iu \bm{m}_{2} \cdot \bm{\mu}}\right) + \sum_{b=2}^{t}\sum_{a=1}^{b-1}h_{a, \bm{m}_1}h_{b, \bm{m}_2}e^{iu(\bm{m}_1 + \bm{m}_2)\cdot \bm{\mu}} \nonumber \\
         =&\left(\sum_{a=1}^{N}h_{a, \bm{m}_1}\right)\left(\sum_{b=1}^{t}h_{b, \bm{m}_2}\right)\frac{e^{iu\bm{m}_2\cdot \bm{\mu}} - e^{iu\left(\bm{m}_1 + \bm{m}_2\right)\cdot \bm{\mu}}}{1 - e^{i\bm{m}_1\cdot \bm{\mu}}}
         + \left(\sum_{b>a=1}^{t}h_{a, \bm{m}_1}h_{b, \bm{m}_2}\right)e^{iu(\bm{m}_1 + \bm{m}_2)\cdot \bm{\mu}}.
         \nonumber \\
    \end{eqnarray}
\end{widetext}
With Eqs.~(\ref{eq.1st}) and (\ref{eq.2nd}), we can construct the fourth-order RDT fluctuations of multiple periods based on the RDTs of one period. With the number of periods $u$ as a variable, the crossing terms include the $e^{iu \bm{m}_2\cdot \bm{\mu}}$ term, the $e^{iu (\bm{m}_1 + \bm{m}_2)\cdot \bm{\mu}}$ term and constant term. Note that the constant term is in Eq.~(\ref{eq.1st}).
And remember that $\sum_{b>a=1}^{t} h_{a, \bm{m}_2}h_{b, \bm{m}_1}$ also drives the same resonance $\bm{m}_1 + \bm{m}_2$.
So there is also the $e^{iu \bm{m}_1\cdot \bm{\mu}}$ term.
Moreover, some fourth-order RDTs consist of more than one pair of crossing terms.
For example, $h_{20110}$ is contributed by $h_{30000}h_{01110}$, $h_{21000}h_{10110}$ and $h_{10200}h_{10020}$ \cite{SLS97}, and there are 8 terms in its fluctuation.

\begin{figure}
    \includegraphics[width=\linewidth]{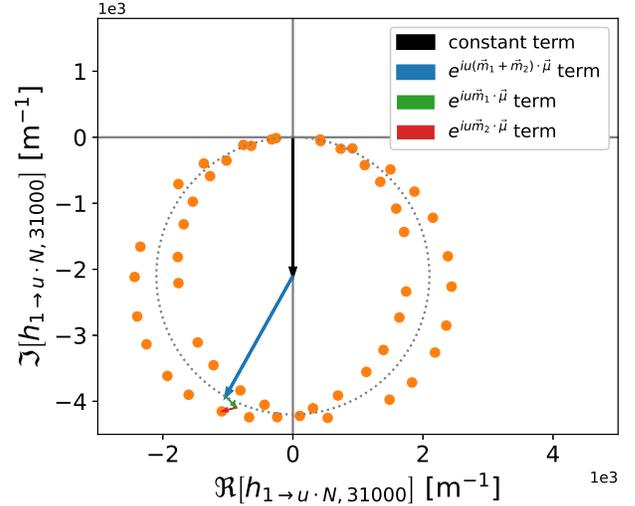}
    \caption{\label{fig:h31000}The values of $h_{1\rightarrow u \cdot N, 31000}$ of the second 6BA lattice contributed by four terms, with $u$ varying from $1$ to $50$.}
\end{figure}

The fourth-order RDT fluctuations described by Eqs.~(\ref{eq.1st}) and (\ref{eq.2nd}) are very complicated.
Here we focus on Eq.~(\ref{eq.1st}) and show an example of the $h_{31000}$ fluctuation of the second 6BA lattice.
The RDT $h_{31000}$ is contributed by $h_{30000}$ and $h_{12000}$ through cross-talk.
Let $\bm{m}_1 = (3, 0)$, $\bm{m}_2 = (-1, 0)$.
The values of $h_{1\rightarrow u\cdot N, 31000}$ of the second 6BA, with $u$ varying from 1 to 50, are shown in Fig. \ref{fig:h31000}.
The constant term and the $e^{iu (\bm{m}_1 + \bm{m}_2)\cdot \bm{\mu}}$ term form a circle passing the origin in the complex plane, so does the fluctuation contributed by octupoles.
According to Sec.~\ref{sec:II}, the coefficient of the $e^{iu \bm{m}_2\cdot \bm{\mu}}$ in Eq.~(\ref{eq.1st}), i.e. $\frac{\sum_{a=1}^{N}h_{a, \bm{m}_1}}{1-e^{i\bm{m}_1\cdot \bm{\mu}}}\frac{\sum_{b=1}^{N}h_{b, \bm{m}_2}}{1-e^{i\bm{m}_2\cdot \bm{\mu}}}$, is exactly the product of the constant terms of the fluctuations of the third-order RDTs $h_{\bm{m}_1}$ and $h_{\bm{m}_2}$.
Generally the third-order RDTs are significantly smaller than the fourth-order RDTs, and then the coefficients of the $e^{iu \bm{m}_1\cdot \bm{\mu}}$ and $e^{iu \bm{m}_2\cdot \bm{\mu}}$ terms are smaller than that of $e^{iu (\bm{m}_1 + \bm{m}_2)\cdot \bm{\mu}}$ term.
As shown in Fig.~\ref{fig:h31000}, the dots are distributed around the dashed circle.
Reducing the third-order RDT fluctuations brings these dots closer to the dashed circle. Moreover, reducing the fourth-order RDT fluctuations leads to smaller radius of the circle.

\bibliography{references}

\end{document}